\documentclass{article}
\usepackage{spconf,amsmath,graphicx}

\title{ECG Artifact Removal from Single-Channel Surface EMG Using Fully Convolutional Networks}
\name{Kuan-Chen Wang$^{1}$ \qquad Kai-Chun Liu$^{2}$ \qquad Sheng-Yu Peng$^{3}$ \qquad Yu Tsao$^{2}$}
\address{
$^{1}$Graduate Institute of Communication Engineering, National Taiwan University, Taiwan\\
$^{2}$Research Center for Information Technology Innovation, Academia Sinica, Taiwan\\
$^{3}$ Department of Electrical Engineering, National Taiwan University of Science and Technology, Taiwan\\
\normalsize{r10942076@ntu.edu.tw, t22302856@citi.sinica.edu.tw, sypeng@mail.ntust.edu.tw, yu.tsao@citi.sinica.edu.tw}
    }
\begin{document}
\ninept
\maketitle
\begin{abstract}
Electrocardiogram (ECG) artifact contamination often occurs in surface electromyography (sEMG) applications when the measured muscles are in proximity to the heart. Previous studies have developed and proposed various methods, such as high-pass filtering, template subtraction and so forth. However, these methods remain limited by the requirement of reference signals and distortion of original sEMG. This study proposed a novel denoising method to eliminate ECG artifacts from the single-channel sEMG signals using fully convolutional networks (FCN). The proposed method adopts a denoise autoencoder structure and powerful nonlinear mapping capability of neural networks for sEMG denoising. We compared the proposed approach with conventional approaches, including high-pass filters and template subtraction, on open datasets called the Non-Invasive Adaptive Prosthetics database and MIT-BIH normal sinus rhythm database. The experimental results demonstrate that the FCN outperforms conventional methods in sEMG reconstruction quality under a wide range of signal-to-noise ratio inputs.
\end{abstract}
\vspace{-1em}
\section{Introduction}
\label{sec:intro}
Surface electromyography (sEMG) noninvasively measures the activation potentials of human muscles by attaching electrodes to the skin. sEMG has been widely adopted in certain applications such as in rehabilitation~\cite{engdahl2015surveying}, stress monitoring~\cite{wijsman2013wearable}, neuromuscular system investigation~\cite{tang2018novel}, and prosthesis control~\cite{ma2014hand}. During the data collection, sEMG would be contaminated by the electrocardiogram (ECG) if the measured muscles are in proximity to the heart~\cite{rissen2000surface,zhou2006eliminating}. ECG contamination distorts the amplitude and frequency spectrum of sEMG, which may further deteriorate the effects of sEMG applications or hinder the determination of relevant information in sEMG signals. Hence, it is crucial to develop denoising techniques and discard ECG artifacts from the sEMG signals.

The general frequency bands of sEMG (10-500 Hz) and ECG (0-100 Hz) partially overlap~\cite{winter2009biomechanics}. This triggers difficulties in discarding ECG artifacts without deteriorating the sEMG quality. To address this problem, various methods have been developed, including high-pass filters (HP) with different cutoff frequencies, template subtraction (TS), adaptive filter (AF) and independent component analysis (ICA). Although these methods have been comprehensively compared and analyzed in previous studies~\cite{xu2020comparative,drake2006elimination}, they remain limited. For instance, applying high-pass filters would eliminate the low-frequency part of sEMG signals; TS works effectively because sEMG is assumed to be a zero-mean Gaussian distribution that is not fully satisfied in an actual environment. Other denoising methods require the use of reference signals. For example, AF requires an additional clean ECG reference~\cite{marque2005adaptive}, and ICA is more suitable for applications with multiple sEMG channels~\cite{willigenburg2012removing}.

Recently, neural networks (NNs) have been widely applied in the development of signal enhancement and denoising approaches in different applications owing to their powerful nonlinear mapping capability, such as acoustic signals~\cite{lu2013speech,FCN_fu2017raw}, ECG~\cite{antczak2018deep,chiang2019noise}, and EEG~\cite{li2015feature,zhang2021eegdenoisenet} noise removal. In these studies, different deep-learning models were developed for signal enhancement. Some commonly adopted models are the multilayer perceptron (MLP)~\cite{lu2013speech,zhang2021eegdenoisenet}, convolutional neural networks (CNNs)~\cite{zhang2021eegdenoisenet}, fully convolutional neural networks (FCNs)~\cite{FCN_fu2017raw, chiang2019noise,FCN_tseng2020study}, and long short-term memory models~\cite{zhang2021eegdenoisenet,LSTM_weninger2015speech}. These NN-based signal enhancement methods have achieved extraordinary results in improving signal quality when compared with conventional denoising methods.

Although previous research has utilized NN to process sEMG signals for other classification tasks (e.g., gesture classification~\cite{allard2016convolutional,wei2019multi} or noise-type identification~\cite{machado2021deep,tosin2021actor}), few studies have explored the feasibility of NN for contamination removal in sEMG~\cite{kale2009intelligent}. Therefore, this study adopted an NN and proposed an FCN-based denoising method to eliminate ECG artifacts from sEMG signals. FCN can handle single-channel sEMGs without other reference signals, e.g., other sEMG channels or an ECG channel. Furthermore, the proposed approach can directly process raw sEMG signals without further data transformation, such as short-time Fourier transform (STFT), which reduces the computational complexity and facilitate near-real-time applications. The experimental results indicate that the FCN-based denoising approach exhibits a better filter ability than conventional methods in the removal of ECG artifacts from sEMG signals. Owing to these advantages, the proposed method may be a suitable choice for eliminating ECG contamination in sEMG.

\vspace{-1em}
\section{Related work}
\label{sec:related}
\subsection{Conventional ECG artifact removal methods}
\label{ssec:ECG removal}
Several methods have been developed to suppress ECG artifacts in sEMG, including the HP, TS, AF and ICA~\cite{xu2020comparative,drake2006elimination,petersen2020removing}. HP directly eliminates the low-frequency part of noisy sEMG signals, where most ECG frequency components exist. TS is based on the quasiperiodic properties of ECG and the assumption that the sEMG has a zero-mean Gaussian distribution. It attempts to eliminate the ECG waveforms from the noisy sEMG signals in the time domain. To develop ECG waveform templates, ECG waveform detection is initially applied to noisy sEMG signals. Subsequently, the ECG templates are acquired with proper filtering or ECG waveform averaging~\cite{xu2020comparative,junior2019template}. Since HP and TS are both applicable for single-channel sEMG, they were implemented in the comparison with the proposed method. A recent study demonstrated that a 4th-order Butterworth filter is effective in ECG artifacts removal~\cite{drake2006elimination}; hence, this study adopted this filter type with a cutoff frequency of 40 Hz, which outperformed other cutoff frequencies on our data. Moreover, the TS method in this study was followed using the HP method to obtain optimal results. AF and ICA are not suitable for single-channel sEMG and are thus excluded from this research. 

\vspace{-1em}
\subsection{NN-based denoise autoencoder}
\label{secc:DDAE}
A denoise autoencoder (DAE) is defined as an autoencoder used to reconstruct clean signals from noisy signals~\cite{vincent2008extracting}. DAE is a machine-learning model that comprises encoder and decoder parts. The encoder transforms the noise-corrupted input $\tilde{x}$ into a hidden representation, and then the decoder maps this hidden representation to the reconstructed signal $\hat x$, which attempts to approach the ground truth $x$ using certain criteria such as the L2 loss. Various studies have verified that the DAE is effective for noise reduction in speech~\cite{lu2013speech}, medical images~\cite{gondara2016medical}, and biomedical signals~\cite{chiang2019noise,li2015feature}. 

The FCN was derived from the CNN model by discarding dense layers in the CNN~\cite{long2015fully}. The model comprises convolutional layers, activation functions, and max-pooling layers. The convolutional layers contain filters to extract feature maps. The max-pooling layers further downsize the feature map by selecting the local maximum values. In addition, the modification of a CNN to an FCN has certain advantages. First, fewer parameters are required in the FCN; hence, the computational complexity can be minimized. Second, the spatial or temporal information of the input data can be better preserved by the mechanism of convolutional layers than by dense layers~\cite{FCN_fu2017raw}. Finally, the fact that the input data size can be unfixed is beneficial to process time-sequence data, whose data length often varies. Owing to these characteristics, this study adopted the FCN model to handle ECG contamination in sEMG signals. The max-pooling layer was not utilized in the proposed method because it may trigger excessive information loss in the detailed structure of the input signals~\cite{chiang2019noise,taigman2014deepface}.

\vspace{-1em}
\section{Materials and methods}
\label{sec:M&M}
In this section, we comprehensively describe the experimental materials, proposed method and evaluation criteria.
\subsection{Open access sEMG and ECG database}
\label{ssec:dataset}
The Non-Invasive Adaptive Prosthetics (NINAPro) database was utilized in this study~\cite{atzori2014electromyography}. In total, 12 channels of sEMG were measured using active wireless electrodes on the upper arm, with a sampling rate of 2 kHz. There were three subsets of data (DB1, DB2, and DB3) in the NINAPro database, and the DB2 with sEMG data collected from 40 intact subjects was selected in this study, as it contained the most subjects. DB2 included three sessions (Exercises 1, 2, and 3), and this study adopted the first two sessions. There were 17 and 22 types of movements performed by each subject during Exercises 1 and 2, respectively. Every type of movement was repeated six times consecutively for 5 s and then followed by 3-s rest. Previous studies have regarded this database as a clean sEMG source after proper filtering~\cite{machado2021deep}. 

For ECG artifacts, this study adopted the MIT-BIH normal sinus rhythm database (NSRD) from the Physionet data bank~\cite{goldberger2000physiobank}. There were 2 ECG channels collected from 18 healthy individuals, and the ECG sampling rate was 128 Hz. Previous studies have also used this database as the ECG noise in sEMG for noise-type detection~\cite{machado2021deep}.

\subsection{Data preprocessing and preparation}
\label{ssec:Data method}
As mentioned in Section 2, this study adopted the sEMG data of Exercises 1 and 2 in the NINAPro DB2 from 40 subjects. The sEMG data were filtered by a 4th-order Butterworth bandpass filter with cutoff frequencies of 20 and 500 Hz and downsampled to 1 kHz. Subsequently, all sEMG signals were normalized by dividing the maximum absolute value and separated into 60 s per segmentation. The ECG data in Channel 1 from the MIT-BIH NSRD were filtered by a 3rd-order Butterworth high-pass and low-pass filter with cutoff frequencies of 10 Hz and 200 Hz, respectively, to discard possible noise in ECG signals~\cite{xu2020comparative}.

We selected the sEMG segments in Channel 2, Exercise 1, from 30 subjects as the training and validation sets. Among them, the sEMG segments from 25 subjects were used for training, and those from the other four subjects were adopted for validation. For each segments, 5 randomly selected ECG signals from 14 subjects in the MIT-BIH NSRD were considered ECG artifacts and superimposed onto the clean sEMG segments at 6 signal-to-noise ratios (SNRs) (-5, -7, -9, -11, -13, and -15 dB). The SNR is defined by Equation~\ref{eq:SNRin}.

To ensure the effectiveness of the proposed methods, mismatch conditions were designed between the training and testing sets. The sEMG segments in Channels 9 to 12, Exercise 2, from subjects 31 to 40 were selected as the testing data. The ECG from the remaining four subjects (19090, 19093, 19140, and 19830) were adopted as noise and subsequently superimposed on the clean sEMG segments at eight SNRs (-14-0 dB with a step of 2 dB). Accordingly, 35300 and 6400 segments of noisy sEMG data were prepared for training and testing, respectively.

\vspace{-1em}
\subsection{Network structure of the FCN-based denoise method}
\label{ssec:FCN structure}
\begin{figure}
    \centering
    \includegraphics[width=\columnwidth]{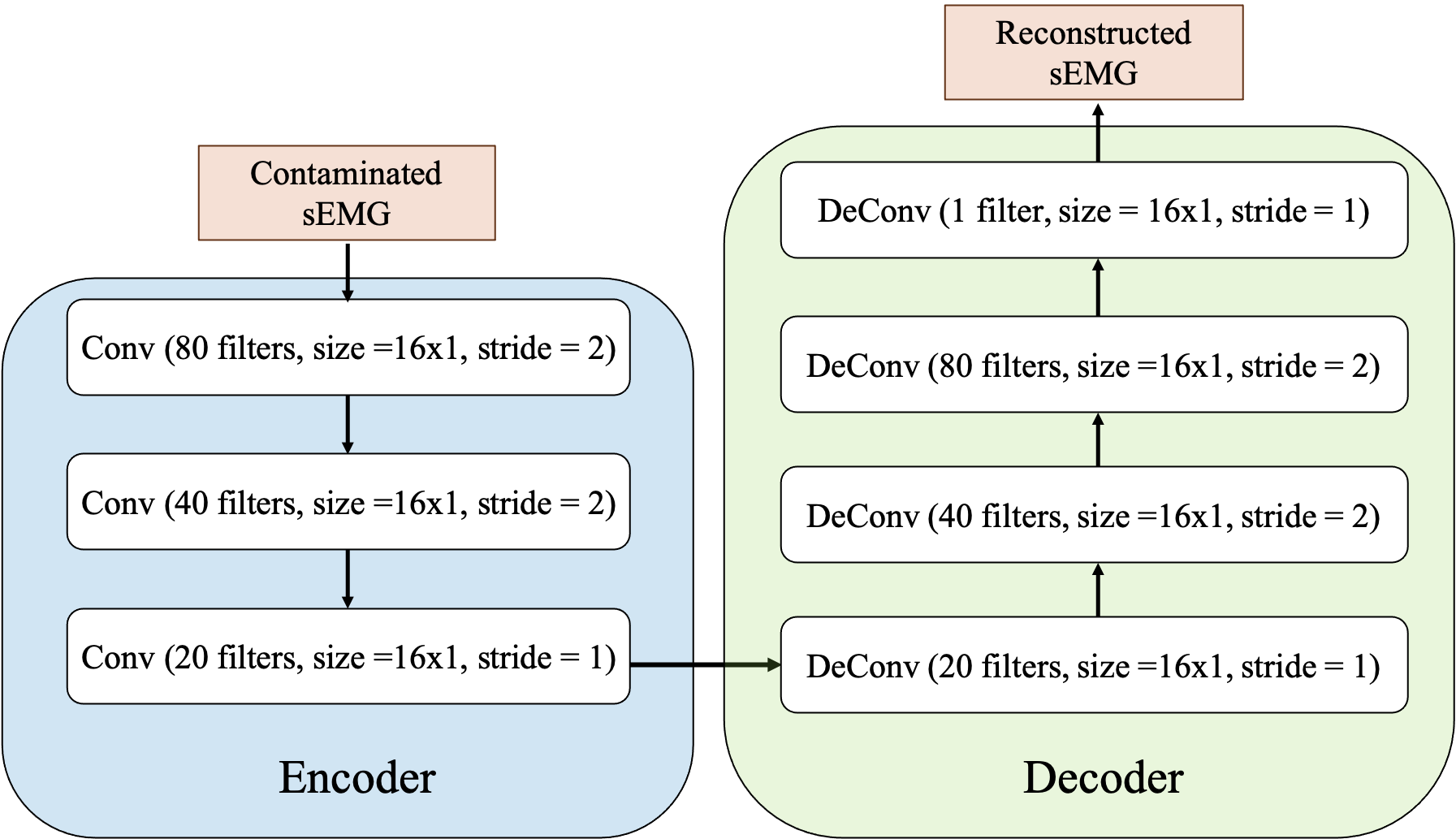}
    \caption{FCN architecture.}
    \label{fig:FCN}
\end{figure}

Fig. \ref{fig:FCN} illustrates the overall network structure of the proposed approach. It comprises two parts: an encoder and a decoder. The encoder with three convolutional layers transformed the noisy sEMG input vector with d $\times$ 1 dimensions into feature maps with (d/2) $\times$ 20, (d/4) $\times$ 40, and (d/4) $\times$ 20 dimensions layer by layer. The input data were compressed to a time length of 1/4 after the first two layers, whose stride was set to 2. The decoder attempts to reconstruct the clean data with d $\times$ 1 dimension  from the encoded feature map. With a symmetric design, the decoder with four deconvolutional layers transformed the (d/4) $\times$ 20 dimension feature map to (d/4) $\times$ 20, (d/2) $\times$ 40, and d $\times$ 80 dimensions, and finally back to a d $\times$ 1-dimensional vector. The filter sizes of all convolutional layers were set to 16 $\times$ 1. Each hidden layers in our model includes an exponential linear unit as the activation functions and a batch normalization layer. No activation function was adopted for the output layer.

\vspace{-1em}
\subsection{Evaluation criteria}
\label{ssec:Evaluation}
This study evaluated the performance of ECG removal methods using two types of criteria: signal reconstruction quality and feature extraction errors. The former category includes a root-mean-square error (RMSE) and SNR improvement (SNR$_{imp}$). RMSE indicates the variance between the reconstructed output and the ground truth, while SNR$_{imp}$ represents the difference between the SNR after noise reduction and the original input signal SNR. These criteria have been widely adopted to evaluate the outcomes of signal-enhancement studies~\cite{chiang2019noise,zhang2021eegdenoisenet}. The calculations were preformed using the following equations:
\begin{equation}
\label{eq:RMSE}
    \text{RMSE}=\sqrt{\frac{\sum_{n=1}^{N}(x[n]-\hat x[n])^2}{N}},
\end{equation}
\begin{equation}
\label{eq:SNRimp}
    \text{SNR}_{imp}=\text{SNR}_{out}-\text{SNR}_{in}.
\end{equation}

SNR$_{out}$ and SNR$_{in}$ denote the SNR values of the output and input sEMG signals, respectively. They are defined as:

\begin{equation}
\label{eq:SNRout}
\text{SNR}_{out}=10\log_{10}{\bigg(\frac{\sum_{n=1}^{N}x[n]^2}{\sum_{n=1}^{N}(x[n]-\hat x[n])^2}\bigg)},
\end{equation}
\begin{equation}
\label{eq:SNRin}
    \text{SNR}_{in} = 10\log_{10}{\bigg(\frac{\sum_{n=1}^{N}x[n]^2}{\sum_{n=1}^{N}(x[n]-\tilde x[n])^2}\bigg)}.
\end{equation}

Better signal reconstruction quality can be represented by smaller RMSE values and larger SNR$_{imp}$ values. 

The other criteria include the RMSE of the average rectified value (ARV) and mean frequency (MF) feature vectors, which are extracted from the sEMG signals. This study adopted these criteria to evaluate the performance of ECG removal methods, as ARV and MF are frequently used features in sEMG applications~\cite{xu2020comparative}. ARV is defined as
\begin{equation}
\label{eq:ARV}
    \text{ARV} = \frac{\sum_{n=1}^{L}|x[n]|}{L}.
\end{equation}
where L denotes the sliding window length. ARV was calculated in a 1-s sliding window without overlap~\cite{xu2020comparative}; hence, L was set to 1000. 

MF describes the power spectrum distribution. In this study, MF is defined as the expected value of the STFT amplitude spectrum between 10 and 500 Hz~\cite{xu2020comparative}. The calculation can be expressed as:
\begin{equation}
\label{eq:MF}
    \text{MF} = \frac{\sum_{n=N_1}^{N_2}f_n\cdot S_n}{\sum_{n=N_1}^{N_2}S_n}.
\end{equation}
where $f_n$ and $S_n$ denote the frequency and amplitude of the sEMG spectrogram, respectively. The sliding window in STFT also has an 1-s window length with no overlap, as did the ARV. Moreover, MF is only calculated in the activation duration, where specific exercises are performed. 

For both the ARV and MF, smaller RMSE values of the extracted feature vectors can ensure the robustness of the performance in applications that employs these sEMG features.

\subsection{Implementation details}
\label{ssec:details}
In this study, the L2 loss and Adam optimizer were applied to update the weights. The learning rate was set at 0.0001. The training was completed when the validation loss stopped dropping after 15 epochs to avoid an overfitting, and the network parameters with the least validation loss were preserved.\footnote{Implementation of the proposed FCN can be acquired in https://github.com/eric-wang135/ECG-removal-from-sEMG-by-FCN}

\vspace{-1em}
\section{Results and discussion}
\label{sec:results}
\begin{figure}
    \centering
    \includegraphics[width=\columnwidth]{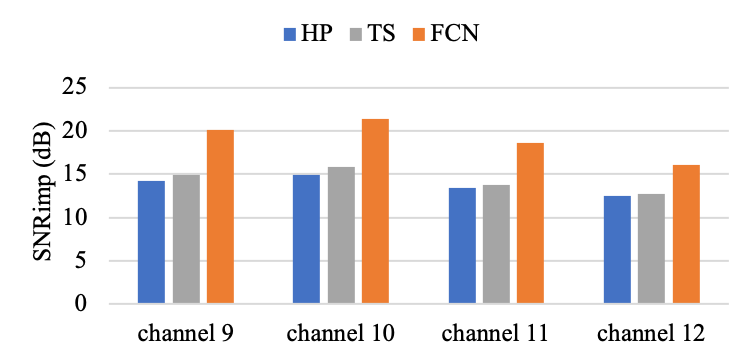}
    \caption{Performance of ECG removal methods evaluated by the SNR$_{imp}$. The average SNR$_{imp}$ values in each sEMG channel are presented in the figure. As illustrated, the FCN outperforms TS and HP in all channels.}
    \label{fig:SNRimp}
\end{figure}

Fig.~\ref{fig:SNRimp} illustrates the overall performance of the ECG artifacts removal methods measured by the SNR$_{imp}$ criterion for four sEMG channels. For every channel, the average SNR$_{imp}$ values of the FCN are higher than those of the other conventional methods by approximately 3.4-5.2 dB. This indicates that, on average, the FCN-based method can render a better reconstruction signal quality than the other methods.

\begin{figure}
    \centering
    \includegraphics[width=\columnwidth]{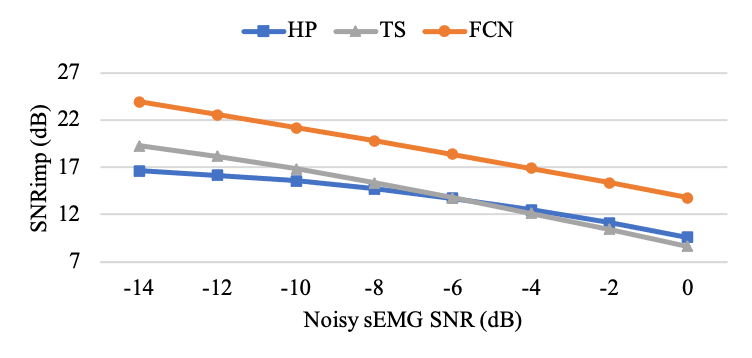}
    \caption{Average SNR$_{imp}$ of sEMG testing data with different SNR inputs. The FCN outperforms TS and HP in all input SNR cases, while the HP and TS curves intersect at -6 dB.}
    \label{fig:SNRimp_crossSNR}
\end{figure}

Fig.~\ref{fig:SNRimp_crossSNR} presents the average SNR$_{imp}$ for the input sEMG with SNR values of -14-0 dB. Evidently, HP and TS are suitable for high and low SNRs, respectively. Compared to conventional approaches, FCN has a more robust performance for all SNR inputs, as the SNR$_{imp}$ of FCN is higher than TS and HP by approximately 4 dB. This further proves that the FCN is a suitable choice for ECG artifacts removal as it maintains the best performance under a broad range of SNR inputs.

\begin{figure}
    \centering
    \includegraphics[width=\columnwidth]{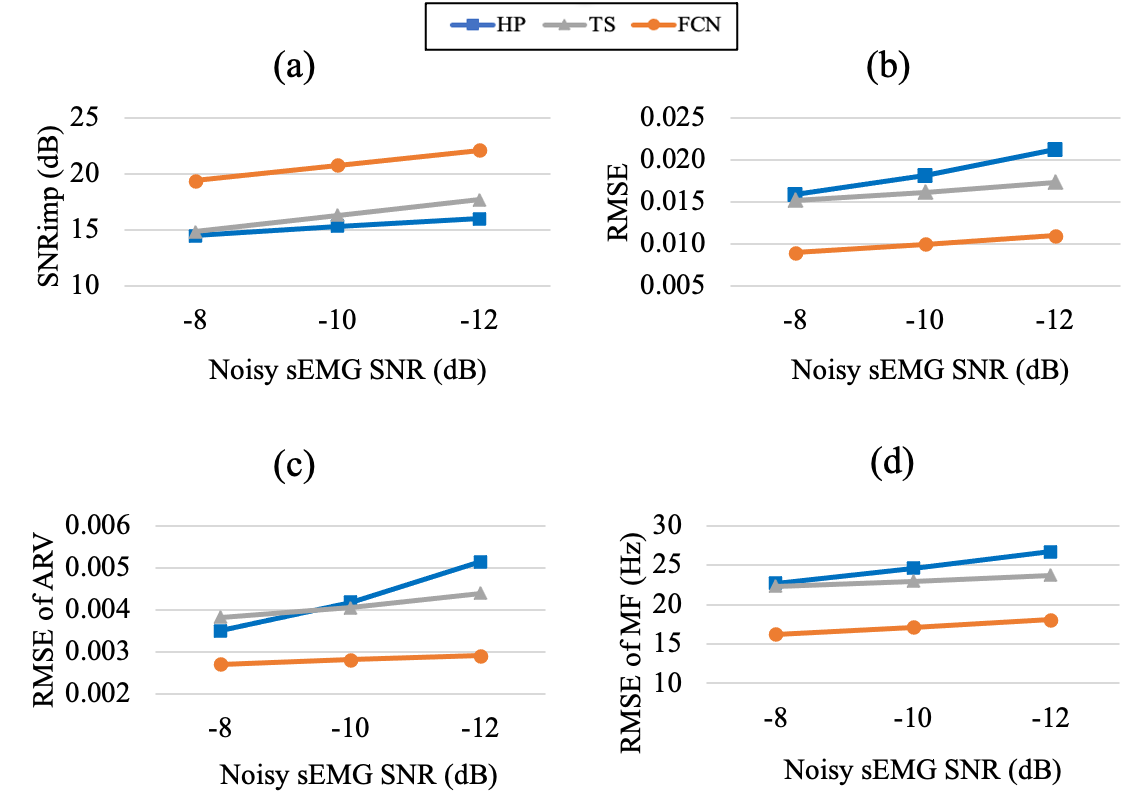}
    \caption{Performance of all methods evaluated by (a) SNR$_{imp}$, (b) RMSE, RMSE of (c) ARV and (d) MF on testing data: Channel 11 sEMG signals contaminated by ECG at SNR = -8, -10, and -12 dB.}
    \label{fig:All_criteria}
\end{figure}

This study also evaluated the performance of the proposed method under a specific scenario. According to previous studies, trunk sEMG signals are often contaminated by ECG at SNR -10 dB~\cite{zhou2006eliminating}, and certain studies have utilized biceps brachii sEMG as simulation data~\cite{xu2020comparative,drake2006elimination}. Hence, this study further applied all the criteria (SNR$_{imp}$, RMSE, and RMSE of ARV and MF) for a comprehensive inspection of ECG artifact removal methods. Fig.~\ref{fig:All_criteria} presents the testing results for Channel 11 sEMG, which were collected from the biceps brachii and contaminated by ECG at SNRs in proximity to -10 dB (i.e., -8, -10, and -12 dB). Under such conditions, the FCN remains the preferred method for ECG artifact removal, according to the applied evaluation metrics.

\begin{figure}[h!]
    \centering
    \includegraphics[width=\columnwidth]{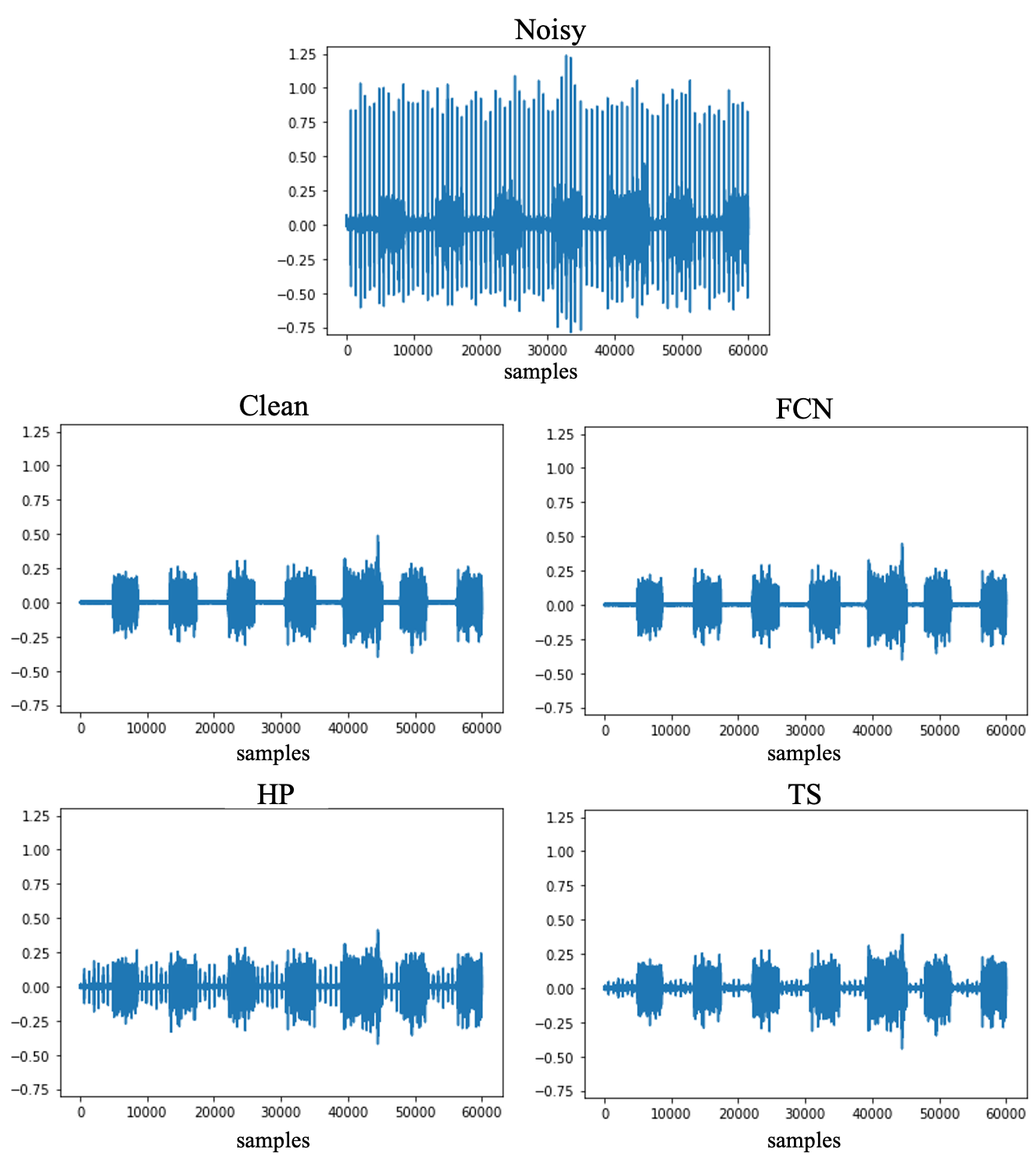}
    \caption{Waveform of the reconstructed sEMG signals by different ECG artifact removal methods. The input noisy sEMG data is Channel 11 sEMG contaminated by the ECG signals of subject 19140 at SNR = -10 dB.}
    \label{fig:waveform}
\end{figure}

Fig.~\ref{fig:waveform} presents the noisy, clean and enhanced sEMG waveforms obtained via the ECG removal methods. The input sEMG data are the Channel 11 sEMG segments contaminated by the ECG signals of subject 19140 at an SNR of -10 dB. Regardless, significant residuals of ECG artifacts emerge in the reconstructed sEMG signals after applying HP (SNR = 4.922 dB) and TS (SNR = 8.710 dB); however, the FCN results are more similar to those of the clean sEMG (SNR = 12.811 dB).

In addition to their noise elimination capability, another important consideration of ECG removal methods is their efficiency. The advantages of the proposed approach is that it can process raw sEMG data directly without any data transformation method, e.g. STFT. In this research, we observed that the FCN model requires significantly less time for filtering than TS. Although FCN cannot outperform HP in terms of time complexity, its remarkable performance and moderate efficiency make it a suitable choice for ECG removal. Several algorithms for accelerating NNs, such as parameters pruning~\cite{karnin1990simple}, can further reduce the time complexity of the FCN. These algorithms could make the FCN more suitable for near-real-time applications.

This study exhibited certain limitations. First, the study focused on eliminating ECG artifacts from a single-channel sEMG. Several other sEMG noise types are yet to be included in this study; however, they may trigger distortion in sEMG signals, similar to ECG. For example, power line interference was discarded by the Hampel filter during the preprocessing stage of the sEMG database. Movement artifacts and white Gaussian noise should also be considered in noise removal. Second, this work solely adopts simulated noisy sEMG data; hence, the effectiveness of the proposed method on actual noisy sEMG data should be verified. Furthermore, several powerful NN models or techniques have been developed in signal-enhancement areas\cite{kim2020t,yi2021eegdnet}; hence, instead of adopting an FCN, it is possible to acquire a better performance for ECG removal with other advanced NN methods.
\vspace{-1.5em}
\section{Conclusion}
\label{sec:conclusion}
This study proposed an FCN-based denoising method for ECG artifacts removal from single-channel sEMG. The proposed FCN technique adopted a denoise autoencoder structure and fully convolutional networks to utilize its powerful nonlinear mapping capability. Moreover, the FCN required no additional reference signal and processed raw sEMG waveforms directly without data transformation, such as STFT. The obtained experimental results indicate that the sEMG denoised by FCN could exhibit a higher average SNR$_{imp}$ than conventional methods (HP and TS), and its optimal performance was maintained under a broad range of SNR inputs. This study also evaluated the feasibility of FCN in specific circumstances of trunk sEMG contamination, and it demonstrated that FCN can outperform other methods in such situations. To the best of our knowledge, this is the first study to apply deep-learning models to ECG artifact removal. In the future, we intend to test this model with zactual contaminated sEMG data. In addition, other sEMG noise types, e.g., power line interference, movement artifacts and white Gaussian noise, would be further considered to develop a robust sEMG noise removal system.

\bibliographystyle{IEEEbib}
{\footnotesize
\bibliography{refs}}

\end{document}